\documentclass[nofootinbib,superscriptaddress,twocolumn]{revtex4-1}
\usepackage{amsmath}
\usepackage{amsfonts}
\usepackage{times}
\usepackage{graphicx}

\usepackage[utf8x]{inputenc}
\usepackage[T1]{fontenc}
\begin{document}

\noindent
http://dx.doi.org/10.1103/PhysRevLett.106.071301

\bigskip

%\title{{\Huge{\ECFTeenSpirit{{ What about Bob?}}}}\\
%\title{Taming nonlocality in theories with deformed Poincar\'e symmetry}
\title{Taming nonlocality in theories with Planck-scale-deformed Lorentz symmetry}

\author{Giovanni AMELINO-CAMELIA}
%\email{Giovanni.Amelino-Camelia@roma1.infn.it}
\affiliation{\footnotesize{Dipartimento di Fisica, Universit\`a di Roma ``La Sapienza", P.le A. Moro 2, 00185 Roma, EU}}
\affiliation{\footnotesize{INFN, Sez.~Roma1, P.le A. Moro 2, 00185 Roma, EU}}
\author{Marco MATASSA}
%\email{Giovanni.Amelino-Camelia@roma1.infn.it}
\affiliation{\footnotesize{Dipartimento di Fisica, Universit\`a di Roma ``La Sapienza", P.le A. Moro 2, 00185 Roma, EU}}
\author{Flavio MERCATI}
%\email{Giovanni.Amelino-Camelia@roma1.infn.it}
\affiliation{\footnotesize{Dipartimento di Fisica, Universit\`a di Roma ``La Sapienza", P.le A. Moro 2, 00185 Roma, EU}}
\affiliation{\footnotesize{INFN, Sez.~Roma1, P.le A. Moro 2, 00185 Roma, EU}}
\author{Giacomo ROSATI}
\affiliation{\footnotesize{Dipartimento di Fisica, Universit\`a di Roma ``La Sapienza", P.le A. Moro 2, 00185 Roma, EU}}
\affiliation{\footnotesize{INFN, Sez.~Roma1, P.le A. Moro 2, 00185 Roma, EU}}

\begin{abstract}
\noindent
We report a general analysis of worldlines  for
 theories with deformed relativistic
symmetries and momentum dependence of the speed of photons.
Our formalization is faithful to
Einstein's program, with
spacetime
points viewed as an abstraction of physical events. The emerging picture imposes the renunciation of the idealization of
absolutely coincident events, but is free from some
pathologies which had been previously conjectured.
\end{abstract}

\maketitle

Over the last decade there has been considerable interest in the quantum-gravity literature
about the possibility~\cite{gacdsr} of deformations of Lorentz symmetry
that would allow the introduction of
 a momentum dependence of the speed of photons
\begin{equation}
v = 1 - \ell \, p
\label{speedphot}
\end{equation}
as a relativistic law, with an observer-independent length parameter $\ell$
usually assumed to be roughly of the order of the Planck length.
This is the most studied possibility for
a ``Doubly-Special Relativity" (DSR)
\cite{gacdsr,kowadsr,leedsrPRL,dsrnature,gacdsr2010review}.
The interest it attracts is
mostly due to associated features that emerge in energy-momentum space,
which find some support in preliminary results obtained within the
Loop Quantum Gravity approach~\cite{leedsrLQG} and in some models based on
spacetime noncommutativity~\cite{jurekDSRnew}.
But the development of this research program
must face the challenge of
several indirect arguments (see, {\it e.g.}, Ref.~\cite{gacIJMPdsrREV,grilloSTDSR} and references therein)
suggesting that a logically consistent formulation of
(\ref{speedphot}) is not possible within  a fully conventional description
of spacetime.

The possibility of novel properties for spacetime
was expected
at the onset~\cite{gacdsr} of DSR research, since the motivation
for the proposal
came from some aspects of the quantum-gravity problem,
which also suggest that there might be some absolute limitations to localizability
of an event.
But the fact it was expected does not make it any less of a challenge: what could replace
the classical points of spacetime?
%It is easy enough to adopt some technical/formal {\it ansatz},
%for example introducing noncommutativity of the coordinates, but what would that mean
%physically? How would that affect the lesson we learned from Einstein concerning the
%fact that spacetime points only acquire physical meaning when they mark
%a physical event?

Those who looked at DSR research from the outside have been understandably rather
puzzled (see, {\it e.g.}, Ref.~\cite{unruh}) about some of the  implications of
renouncing to an ordinary spacetime picture.
In particular, the recent Ref.~\cite{sabinePRL} ventured to make a bold claim:
even without adopting any specific formalization, using only the bare idea
of momentum-dependence of the speed of photons, one could
robustly estimate the nature and size of the nonlocal effects that should
be produced. And, still according to
Ref.~\cite{sabinePRL}, this could be used to constrain $\ell$
to $|\ell| < 10^{-58}m$, {\it i.e.} at a level which
 is 23 orders of magnitude beyond the one of direct experimental bounds based on
the momentum dependence of the speed of photons~\cite{fermiNATURE,dsrFRW}.

The claim reported in Ref.~\cite{sabinePRL} clearly renders even more
 urgent for DSR research to establish what are the actual implications for nonlocality.
We start by observing that the argument presented in Ref.~\cite{sabinePRL}
did not make use of the well-established  results on DSR-deformed boosts,
 but rather relied on assumptions that fail
to be consistently relativistic.
As shown in Fig.~1 the assumptions of  Ref.~\cite{sabinePRL}
amount to adopting {\underline{undeformed}} rules of boost transformation
for the coordinates of the emission points of particles but
{\underline{deformed}}
boost transformations for the velocities of the particles.
Evidently such criteria
 of ``selective applicability" of deformed boosts
 cannot produce a consistently relativistic picture.

\begin{figure}[h!]
\begin{center}
\includegraphics[width=0.33\textwidth]{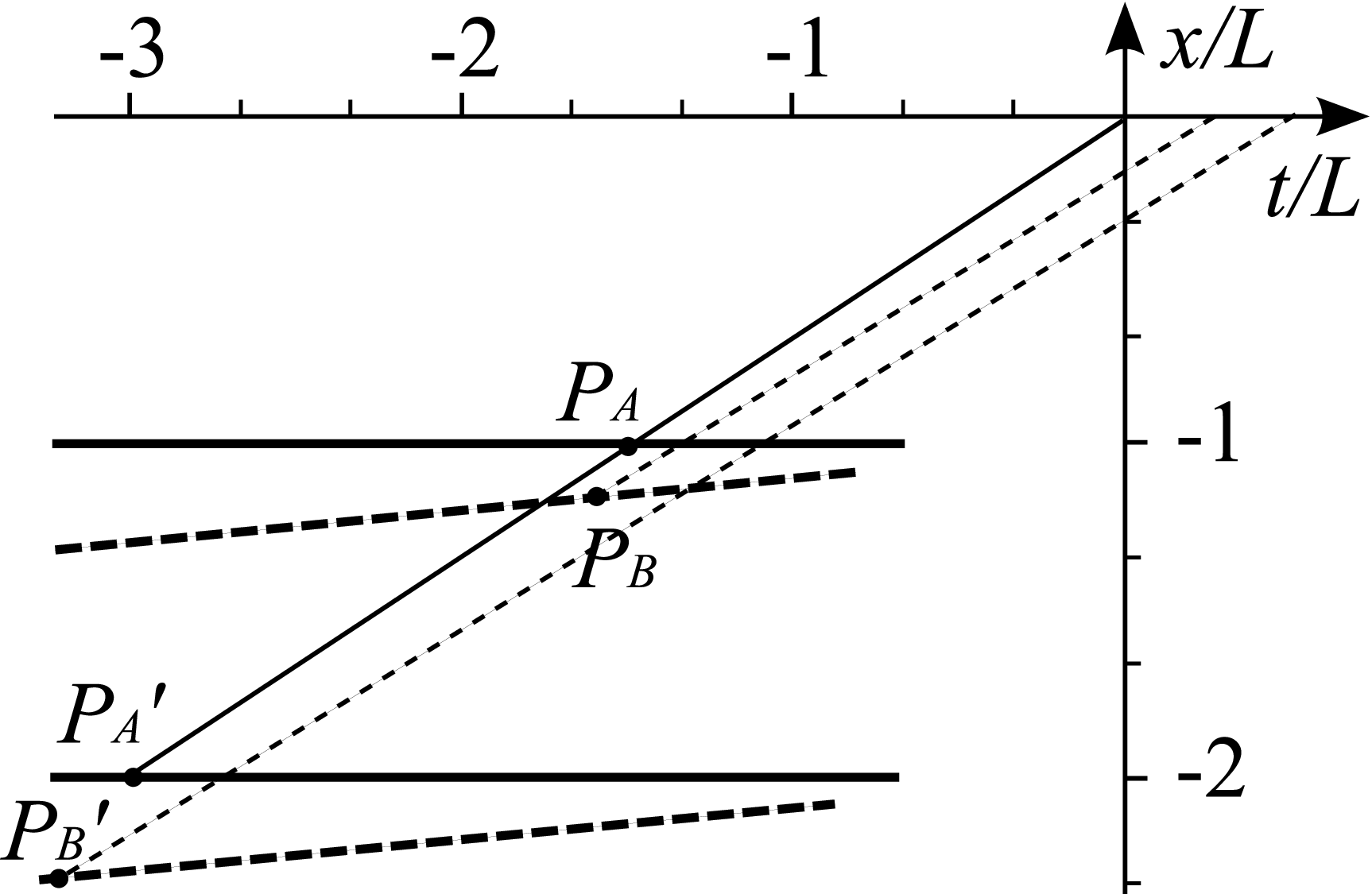}
\end{center}
\caption{In the argument of Ref.~\cite{sabinePRL} a key role is played by
the assumption that
a photon which Alice sees emitted in $P_A$ (from a source at a distance $L$ from Alice)
with  speed $v$ (and momentum $p$)
should be seen by boosted Bob as a photon
emitted from $P_B$, obtained by classical/{\underline{undeformed}} boost
of $P_A$, with speed obtained from the speed $v$ with a {\underline{deformed}} boost.
In this figure we expose the logical inconsistency of such criteria
 of ``selective applicability" of deformed boosts
by allowing for a second photon which according to Alice
also has speed $v$
and is emitted from a point $P_{A'}$ such that the two photons
share the same worldline: a single worldline would be mapped by
a relativistic boost into two wildly different worldlines.}
\end{figure}

The picture proposed in Ref.~\cite{sabinePRL} clearly needed to be revised.
We here report a deductive result of
characterization of the nonlocality produced by DSR boosts. We derive it rigorously
from the formalizations of DSR-deformed boosts that have been proposed in the
DSR literature.
We succeed, where others had failed, primarily as a result of using as guidance
Einstein's insight on the proper characterization of a spacetime point,
to be viewed as the abstraction of an event of crossing of worldlines.
This leads us to a fully relativistic characterization of the concept of locality,
as a concept that pertains the coincidence of events: from a relativistic perspective
the main locality issue concerns whether events that are coincident for one
observer are also coincident for other observers.

We set to $1$ both Planck constant $\hbar$
and the speed-of-light scale $c$ (speed of photons in the low-momentum limit).
The modulus of a spatial 3-vector, with components $W_j$, is denoted
by $W$ ($W^2 \! = \! W_j W^j$).
And we work in leading order in $\ell$,
since (\ref{speedphot}) is assumed~\cite{gacdsr,leedsrLQG} to be valid
only for $p \ll 1/|\ell|$.

We also take on the challenge of a full 3+1-dimensional analysis.
Most of the previous DSR literature, including Ref.~\cite{sabinePRL},
is confined to 1+1-dimensional frameworks, as a way to temper
the complexity of dealing with deformed boosts. It is natural to expect
that the core implications for the nonlocality produced by DSR boosts
would be already uncovered in a 1+1-dimensional analysis, but
our ability to characterize transverse boosts is a valuable addition,
and provides further evidence of the robustness of the approach we developed.
Another significant strength of our setup is that it applies to all previously considered  deformations of Lorentz
symmetry compatible with (\ref{speedphot}). Previous DSR studies of (\ref{speedphot})
not only failed to offer an explicit analysis
of worldlines, but were also often assuming a specific {\it ansatz}
 for the formalization of the symmetry deformation.

 The derivation of the worldlines is here achieved
 within a Hamiltonian setup
which was already fruitfully applied~\cite{dash,mignemiHAMILT} to
other DSR scenarios for the introduction of the
second relativistic scale $\ell$, but was not previously implemented for
a DSR description of
the speed law  (\ref{speedphot}) for massless particles.
We start by introducing canonical momenta conjugate
to the coordinates  $x_j$ and $t$:
 $\{ \Pi_j , x_k \} = -\delta_{jk}$, $\{ \Omega , t \} = 1$.
We must then specify a form of the DSR-deformed mass Casimir $\mathcal{C}$,
which will play the role~\cite{dash,mignemiHAMILT} of Hamiltonian.
We have a two-parameter family of $O(\ell)$ possibilities
\begin{equation}
\mathcal{C} = \Omega^2 - \Pi^2 + \ell \left( \gamma_1 \Omega^3 +\gamma_2 \Omega \Pi^2\right) ~,
\label{casimir}
\end{equation}
upon enforcing analyticity of the deformation and invariance under classical space-rotation transformations. The types of deformed boosts that were previously
considered in the DSR literature have the property of being compatible with such
a deformed Casimir, for some corresponding choices of $\gamma_1$,$\gamma_2$.

Hamilton's equations give the conservation of $\Pi_j$ and $\Omega$ along the worldlines
\begin{equation}
\dot{\Pi}_j = \frac{\partial \mathcal{C}}{\partial x_j} = 0 ~,
 ~~~ \dot \Omega = - \frac{\partial \mathcal{C}}{\partial t} = 0 ~,
\end{equation}
where $\dot f \equiv \partial f/ \partial \tau$ and $\tau$ is an auxiliary worldline parameter.

The worldlines can then be obtained observing that
\begin{equation}
\begin{split}
&\dot{x}_j = - \frac{\partial \mathcal{C}}{\partial \Pi_j} ~ \Rightarrow ~ x_j (\tau)
= x_j^{(0)} + ( 2 \Pi_j
- 2 \ell \gamma_2 \Omega \Pi_j) \tau \\
&\dot t = \frac{\partial \mathcal{C}}{\partial \Omega} ~ \Rightarrow ~ t(\tau) = t^{(0)} \!+\!\!\left[2 \Omega + \ell (\gamma_2 \Pi^2 + 3 \gamma_1 \Omega^2) \right] \tau~.
\end{split} \label{HamiltonEquations}
\end{equation}
Eliminating the parameter $\tau$  and imposing the Hamiltonian constraint $\mathcal{C} = 0$ (massless case) one finds that
\begin{equation}
x_j = x_j^{(0)}
+ \frac{\Pi_j}{\Pi}  (t-t^{(0)}) - \ell (\gamma_1 + \gamma_2 ) \Pi_j (t-t^{(0)})~,
\label{worldlines}
\end{equation}
which reproduces (\ref{speedphot}) for $\gamma_1 + \gamma_2 = 1$.
Note that this derivation of worldlines compatible with (\ref{speedphot})
is insensitive to the possibility
of a different DSR description for the canonical momentum $\Pi_j$ and for
the ``momentum" $p_j$, intended as the DSR generalization of the concept
of space-translation charge. Indeed, $\Pi_j$ enters only at order $\ell$ and of course,
since we are working in leading order, we must take $\ell \Pi_j = \ell p_j$
(while the modulus of $p_j$  and $\Pi_j$
may differ~\cite{gacdsr2010review,mignemiNONCANOMOME} at order $\ell$).

We must now enforce
covariance of the worldlines under DSR-deformed boosts.
The form of the correction terms introduced in (\ref{casimir})
suggests that the type of deformed boosts considered in the
DSR literature should be well suited:
\begin{equation}
\! \mathcal{N}_j \! =  \! - t \Pi + x_j \Omega + \ell [\alpha_1  t \Omega \Pi_j +  \alpha_2 \Pi^2 x_j
 + \alpha_3 \Omega^2 x_j + \alpha_4 x_k \Pi^k \Pi_j ]
 \nonumber
\end{equation}
Note that this four-parameter family of $O(\ell)$
deformed boosts,
which enforces compatibility
with undeformed space rotations,
includes, as different
particular cases, all the proposals for deformed boosts that were put forward
in this first decade
of DSR research~\cite{gacdsr,kowadsr,leedsrPRL,dsrnature,gacdsr2010review,kowaMANY}.
The compatibility between boost transformations and form of the Casimir is encoded
in the requirement that the boost charge is conserved
\begin{equation}
\dot{\mathcal{N}}_j = \{ \mathcal{C} ,  \mathcal{N}_j \} = \frac{\partial  \mathcal{C}}{ \partial \Omega}\frac{\partial  \mathcal{N}_j }{ \partial t} - \frac{\partial  \mathcal{C}}{ \partial \Pi^k } \frac{\partial  \mathcal{N}_j}{ \partial x_k}  = 0 ~,
\end{equation}
which straightforwardly leads to the following constraints on
the parameters $\gamma_1 , \gamma_2 , \alpha_1 , \alpha_2 , \alpha_3 , \alpha_4$:
\begin{equation}
2 \alpha_2 + 2 \alpha_4 = \gamma_2~, ~~
2 \alpha_1 + 2 \alpha_3 - 3 \gamma_1 - 2 \gamma_2=0
 ~. \label{ConstraintsOnBoost}
\end{equation}
Combining these
with the requirement $\gamma_1 + \gamma_2 = 1$ derived above,
we finally arrive at a three-parameter family of Hamiltonian/boost pairs
\begin{eqnarray}
&&\mathcal{C}=\Omega ^2-\Pi ^2+\ell  \left(2 \gamma\Omega ^3+\left(1-2\gamma \right)\Omega \text{  }\Pi ^2\right)
%~,
\nonumber
\\
&& \mathcal{N}_j=-t \Pi_j + x_j \Omega + \ell \, \alpha t \Omega \Pi_j
- \ell \, \left(\gamma + \beta -1/2 \right)  x_k \Pi^k \Pi_j
\nonumber \\
&&~~~~~~~~~~+ \ell x_j \left(\,\beta \Pi ^2 +
 \left(1+\gamma-\alpha \right) \Omega ^2 \right)~,
%\label{nresult}
\nonumber
\end{eqnarray}
where $\gamma = \gamma_1/2$, $\alpha = \alpha_1$,  $\beta = \alpha_2$.
For any given choice of $\gamma , \alpha , \beta$ relativistic covariance is
ensured and we have a rigorous Hamiltonian
derivation of worldlines for which the speed law
(\ref{speedphot}) is verified.
We have so far focused on massless particles, but one also easily obtains the worldlines of particles of any mass
%\footnote{For
%the specific class of DSR deformations we are focusing on, the ones compatible with
%the speed law (\ref{speedphot}) for massless particles, the effects of the deformation
%are further suppressed in the regimes where the mass of a particle can be appreciated.
%But our strategy of formalization of DSR scenarios is clearly applicable also
% to other DSR schemes for the introduction of a relativistically invariant length scale %$\ell$,
% and for some DSR scenarios the study of massive particles is a valuable phenomenological
% opportunity~\cite{gacPRL2009}.}
 by enforcing the Hamiltonian constraint $\mathcal{C} = m^2$:
\begin{equation}
x_j = x_j^{(0)} + \frac{\Pi_j}{\sqrt{\Pi^2+m^2}}  (t-t^{(0)}) - \ell \Pi_j  (t-t^{(0)}) ~.
\label{fullworldlines}
\end{equation}
The covariance of these worldlines under undeformed space rotations is manifest.
The covariance under $\gamma, \alpha ,\beta$-deformed boosts, ensured by construction,
can also be verified
 by computing explicitly the action of an infinitesimal
deformed boost with rapidity vector $\xi_j$ ($A' = A + \xi_j \{ A, {\cal N}^j\}$)
\begin{eqnarray}
&& \!\!\!\!\Pi '_j=\Pi _j-\xi _j \Omega -\ell \, \xi _j \left(\beta \Pi ^2
+\left(1+\gamma -\alpha \right) \Omega ^2 \right) \nonumber\\
&& ~~~~~~~-\ell  \left(1/2-\gamma-\beta \right) \xi_k \Pi^k \Pi_j
\label{ptra}\\ &&
\!\!\!\!t'\!=\!t-\!\xi_j x^j - \!\ell  \left(\alpha t \xi _j \Pi^j+2\left(1 \! + \! \gamma
 \! - \! \alpha\right) \Omega   \xi_j x^j\right)
\label{ttra}\\ &&
\!\!\!\!x'_j = x_j - t \xi_j+\ell  \left(\alpha t \Omega \xi_j+2\beta  \xi_k x^k \Pi_j \right) \nonumber \\
&&~~~~~~ - \ell  \left( \gamma + \beta - 1/2 \right) \left(\xi_k \Pi^k x_j + x_k \Pi^k \xi_j \right) \label{xtra}
\end{eqnarray}
Using these
% transformation rules
one easily verifies that when Alice has the particle
on the worldline (\ref{fullworldlines}) Bob sees the particle on the worldline
\begin{equation}
 x_j' = x'^{(0)}_j + \frac{\Pi'_j}{\sqrt{m^2 +\Pi'^2}} (t'-t'^{(0)}) - \ell \, \Pi'_j (t'-t'^{(0)}) ~,
 \nonumber
\end{equation}
consistently with the relativistic nature of our framework.

We are now ready to exploit our technical results for a ``physical" characterization
of the nonlocality produced by DSR boosts.
The observations we shall make on nonlocality apply equally well
to all choices of $\gamma$, $\alpha$, $\beta$.
We notice however that by enforcing
the condition $\alpha \,\! - \,\! \beta \,\! - \,\! \gamma \! \, = \! 1/2$
one has the welcome~\cite{jurekDSRnew,mignemiHAMILT} simplification
of undeformed Poisson brackets among boosts and rotations
(``the Lorentz sector is classical"~\cite{jurekDSRnew,mignemiHAMILT}).
And in particular
for the case $\gamma \! = \! 1/2$,
$\alpha \! = \! 1 $, $\beta \! = \! 0$, on which we focus
for our graphical illustrations, the laws
of transformation take a noticeably simple form:
%\begin{eqnarray}
%\tilde{\mathcal{N}}_j = -t \Pi _j+x_j \Omega +\ell  \left(\alpha _1 t \Omega  \Pi _j + \Omega %^2 x_j / 2\right)
% ~,
% \label{boostee}
%\end{eqnarray}
%
%The worldlines are the same independently of the choice of $\gamma$, $\alpha_1$ and %$\alpha_2$,
%but of course one needs to specify $\alpha_1$ and $\alpha_2$ in order to give a
%prescription for the relativistic map between two inertial observers with a relative boost.
%Let us for example consider the case
%an infinitesimal Boost generated by $\tilde{\mathcal{N}}_j$ acts on the coordinates of points
%on a worldline and on the canonical momentum $\Pi_j$ as follows,
\begin{eqnarray}
&& \!\!\!\!\Pi '_j=\Pi _j-\xi _j \left( \Omega + \ell \Omega ^2/2 \right)  \label{ptrab}\\
&&
\!\!\!\!t'\!=\!t-\!\xi_j x^j - \!\ell  \left( t \xi _j \Pi^j+ \Omega   \xi_j x^j\right)
 \label{ttrab}\\ &&
\!\!\!\!x'_j = x_j - \left(1- \ell \Omega \right) t \xi_j   \label{xtrab}
\end{eqnarray}
This case preserves
much of the simplicity of classical boosts for what concerns  boosts
acting transversely to the direction of motion. We do not expect
anything objectively pathological in the richer structure that other
choices of  $\gamma$, $\alpha$, $\beta$ produce (see (\ref{ttra})-(\ref{xtra}))
 for such transverse boosts. But it is nonetheless noteworthy
 that there are candidates for the DSR deformed boosts that
 have properties as simple as codified in (\ref{ttrab})-(\ref{xtrab}).
 In what follows we shall not offer any additional comments on
 transverse boosts (and our figures focus on boosts along the
 direction of motion). But it is easy to verify
using (\ref{ttra})-(\ref{xtra})
(and even easier using (\ref{ttrab})-(\ref{xtrab}))
that boosts acting  transversely to
the direction of motion lead to features of nonlocality that are
of the same magnitude and qualitative type as the ones we visualize for boosts along
the direction of motion.

Let us now move on to reconsidering the issues raised in our Fig.~1,
and the shortcomings of the analysis reported in Ref.~\cite{sabinePRL}.
Having  managed to derive constructively quantitative formulas for
the action of the deformed boosts advocated in the DSR literature,
we can now more definitely observe that the assumptions made for
the analysis reported in Ref.~\cite{sabinePRL}
are inconsistent with the fact, here shown in Eqs.~(\ref{ttra})-(\ref{xtra}),
that the deformed boosts still act,
like ordinary Lorentz boosts, in way that is homogeneous
in the coordinates.
A boost connects two observers with the same origin of their reference frames
%(which is expressed more relativistically observing that they must be both local to the
%same event)
and, as shown in Fig.~2, the differences between DSR-deformed boosts
and classical boosts are minute for points that are close to the common
origin of the two relevant reference frames, but gradually grow with
distance from that origin.

As shown by two of the worldlines in Fig.~3, when an observer Alice is local
to a coincidence of
events (the  violet and a red photon simultaneously crossing Alice's worldline)
all observers that are purely boosted with respect to Alice, and therefore share
her origin,  also describe those two events
as coincident.
This in particular addresses the ``box problem"
 raised in Ref.~\cite{sabinePRL},
which concerned the possibility of a loss
 of objectivity of coincidences of events
as witnessed by local observers: we have found that, at least in
leading order in $\ell$ and $\xi$, in the DSR
framework ``locality", a coincidence of
 events,
preserves its objectivity if assessed by local observers.

\begin{figure}[h!]
\begin{center}
\includegraphics[width=0.44\textwidth]{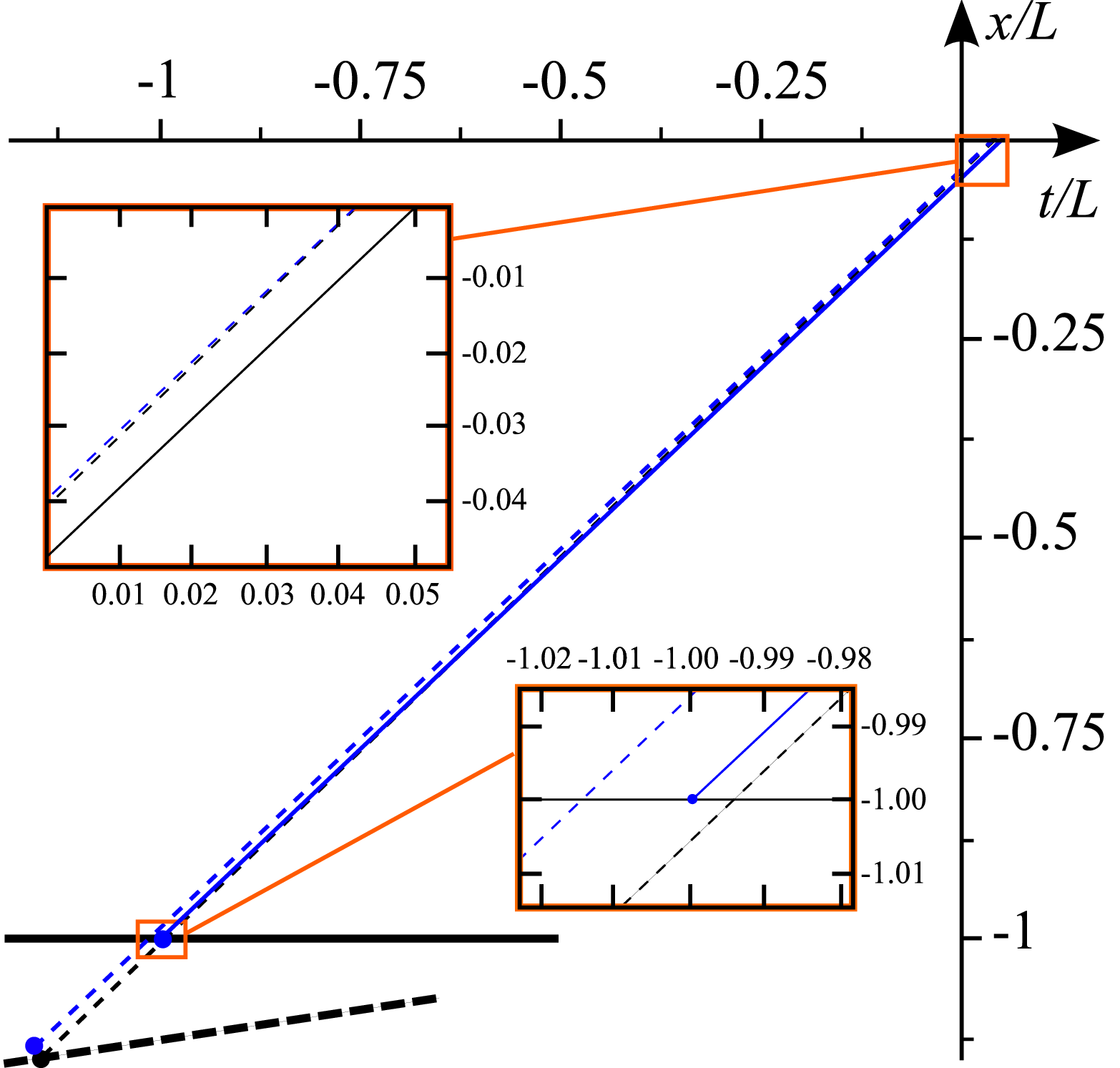}
\end{center}
\caption{We here show a hard-photon worldline
%(missing the origin by a fair amount)
as seen by Alice (solid bue), by DSR-boosted Bob (dashed blue)
and by classically-boosted Bob (dashed-black).
 In spite of assuming (for visibility)
the unrealistically huge $\Pi \! = \! 0.05/\ell$, $\xi \! = \! 0.15$,
the difference between DSR boosts and undeformed boosts
is minute near the origin.
But according
  to Bob's coordinates the emission of the hard particle appears to occur slightly off the
  (thick) worldline of the source.}
\end{figure}

\begin{figure}[h!]
\begin{center}
\includegraphics[width=0.46\textwidth]{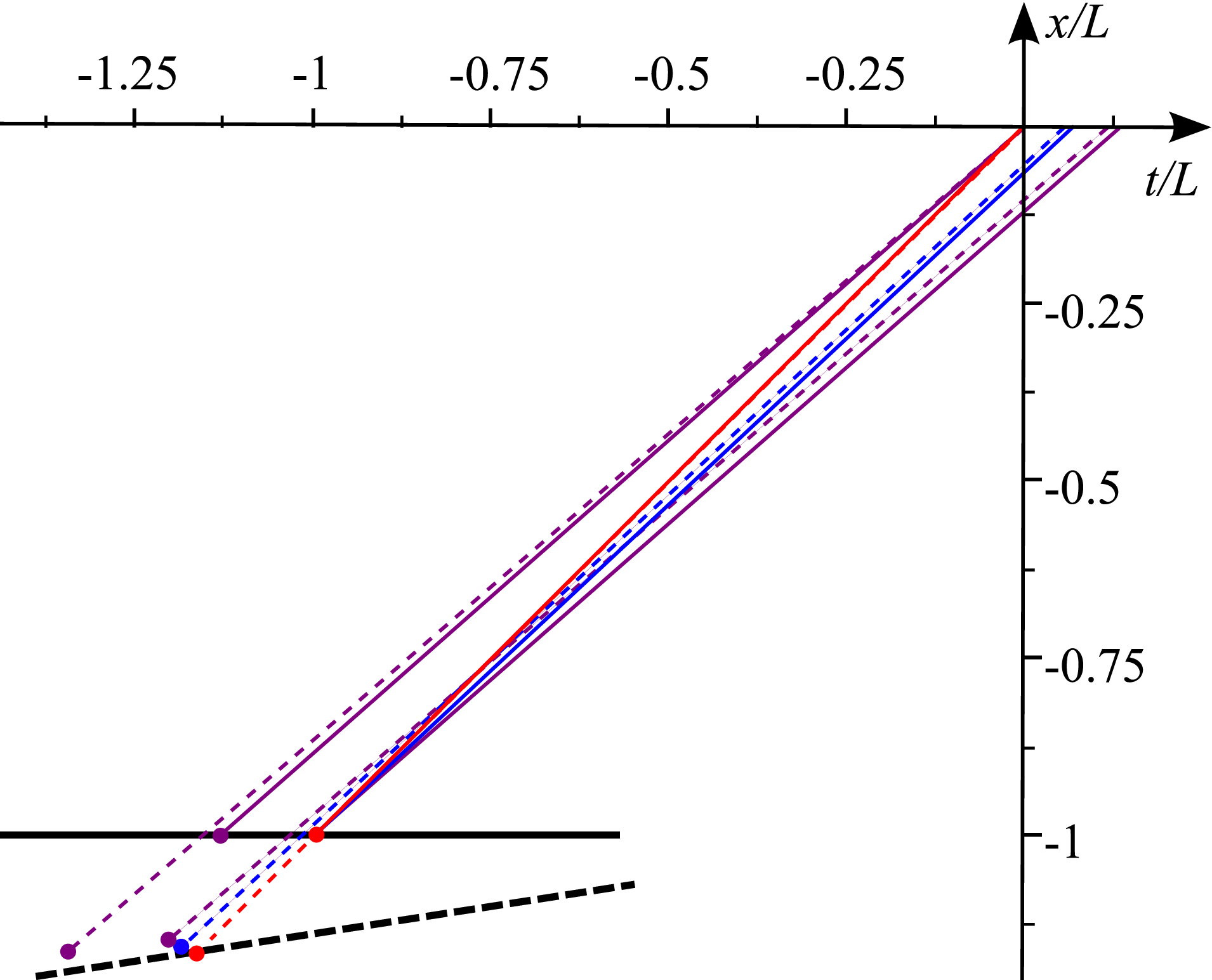}
\end{center}
\caption{A case with two hard (violet) worldlines, with
momentum $\Pi_{v} = 0.13/\ell$, a ``semi-hard" (blue) worldlline
with momentum $\Pi_{b}=\Pi_{v}/2$,
and a ultrasoft worldline (red, with $\Pi_r \ll 1/\ell$).
According to Alice (whose lines are solid, while boosted Bob has dashed lines)
three of the worldlines give
a distant coincidence of events,
while two of the worldlines cross in the origin.}
\end{figure}

The element of nonlocality that is actually produced by
DSR-deformed boosts is seen by focusing on the ``burst" of three photon worldlines
also shown in Fig.~3,
whose crossings establish a coincidence of events for Alice far from her origin,
an aspect of locality encoded in a ``distant coincidence of events".
The objectivity of such distant coincidences of events is partly spoiled by the
DSR deformation: the coincidence is only approximately present
in the coordinates of an observer boosted with respect to Alice.
But we stress that
in Figs.~2 and 3
we used, for visibility, gigantically unrealistic values of photon momentum
(up to $\sim \! 0.1/\ell$): it should nonetheless be noticed that
even distant coincidence is objective
 up to a very good approximation, if indeed,
as assumed in the DSR literature\cite{gacdsr,kowadsr,leedsrPRL,dsrnature,gacdsr2010review},
the observer-independent length scale $\ell$ is as small as the Planck length
($\sim \! 10^{-35}m$). On terrestrial scales one might imagine hypothetically
to observe a certain particle decay with two  laboratories,
with a large relative boost of, say, $\xi \! \sim \! 10^{-5}$,
with idealized absolute accuracy in tracking   back
to the decay
region  the worldlines of two particles that are the decay products.
As one easily checks from (\ref{ttra})-(\ref{xtra}), the
 peculiar sort of nonlocality we uncovered is of size $\xi \ell L \Pi$.
Therefore even if the distance $L$ between the decay region and the observers
is of, say,  $10^4 m$, and the decay products have momenta of, say, $100 GeV$,
one ends up with an apparent nonlocality of the decay region which is only
of $\sim 10^{-19}m$.
% (if indeed $\ell \sim 10^{-35}m$)

Another interesting case
is the one of a typical observation of a gamma-ray burst, with $GeV$
particles that travel for, say, $10^{17}s$ before reaching our telescopes.
For two telescopes
with a relative boost of $\xi \sim 10^{-4}$
the loss of coincidence of events at the source is $\sim 100 m$,
well below the sharpness we are able to attribute~\cite{fermiNATURE} to the location of a gamma-ray burst.

We should stress that actually, in light of the results we obtained, in such a DSR
framework
two relatively boosted observers should not dwell about distant coincidences, but rather
express all observables in terms of local measurements (which is anyway
what should  be done in a relativistic theory). For example,
for the burst of three
photons shown in Fig.~3 the momentum dependence of the speed of photons
is objectively manifest (manifest both for Alice and Bob)
in the linear correlation between arrival times and momentum of the photons.

Also insightful is the comparison of the loss of objectivity of coincidences
of distant events, which we uncovered here for DSR boosts,
with the loss of objectivity of simultaneity that was required by the
replacement of Galileian boosts with Lorentz boosts.
With absolute time of course any statement of simultaneity was objective.
With the introduction of Lorentz boosts, which are obtained deforming Galileian boosts,
simultaneity is no longer objective in general, but it remains objective
for events occurring at the same spatial position, whether or not that
spatial position is where the observer is located (the origin).
With one more step of deformation of boosts, the DSR proposal,
the realm of objectivity of simultaneity is farther reduced:
simultaneity of events is only objective if the events are coincident
according to a local observer, and this is only manifest in the coordinate
systems of other observers that
are also local to the coincidence of events.

Amusingly it appears
that the possibility of coincident events
was cumbersome already for Einstein, as shown by
a footnote in the famous 1905 paper~\cite{einstein}:\\
$~~~~~~~~${\small  ``{\it We shall not discuss here the imprecision inherent in}}\\
 $~~~~~~~~${\small {\it the concept of
simultaneity of two events taking place}}\\
 $~~~~~~~~${\small {\it  at (approximately) the same
location, which can be}}\\
 $~~~~~~~~${\small {\it  removed only by abstraction.}"}\\
We conjecture that the proper description of the quantum-gravity realm,
whether or not there will be a role for DSR concepts,
will impose the renunciation of the idealization of the possibility of exact and absolute coincidence of events.

\end{document}